\documentclass[numberedappendix,twocolumn,twocolappendix]{openjournal}

\usepackage{latexsym}
\usepackage{graphicx}
\usepackage{amssymb}
\usepackage{longtable}
\usepackage{epsf}
\usepackage{amsmath}
\usepackage{graphicx}
\usepackage{gensymb}

\usepackage[backref=page]{hyperref}
\hypersetup{colorlinks=true,linkcolor=blue,citecolor=blue,filecolor=blue,urlcolor=blue}
\usepackage{cleveref}
\usepackage{bm}		% Bold maths symbols, including upright Greek
\usepackage{ulem}
\usepackage{orcidlink}

\usepackage{natbib}

\newcommand{\lcdm}{\xspace{\ensuremath{\Lambda\mathrm{CDM}}}\xspace}
\newcommand{\wowacdm}{\xspace{\ensuremath{w_0 w_a\mathrm{CDM}}}\xspace}

\newcommand{\Msun}{\xspace{$\mathrm{M}_{\odot}$}\xspace}
\newcommand{\hMsun}{\xspace{$h^{-1}\mathrm{M}_{\odot}$}\xspace}

\newcommand{\hgpc}{\xspace{$h^{-1}\mathrm{Gpc}$}\xspace}

\newcommand{\beq}{\begin{eqnarray}}
\newcommand{\eeq}{\end{eqnarray}}
\newcommand{\ben}{\begin{enumerate}}
\newcommand{\een}{\end{enumerate}}
\newcommand{\bit}{\begin{itemize}}
\newcommand{\eit}{\end{itemize}}

\newcommand{\Lbox}{L_{\rm box}}
\newcommand{\mptcl}{m_{\rm ptcl}}

\usepackage{xspace}

\begin{document}

\title{Illuminating the Physics of Dark Energy with the Discovery Simulations\vspace{-1.3cm}}
\shorttitle{Discovery Simulations}

\author{Gillian D. Beltz-Mohrmann$^{1,\star}$\orcidlink{0000-0002-4392-8920}, Adrian Pope$^{2}$\orcidlink{0000-0003-2265-5262}, Alex Alarcon$^{1,4}$\orcidlink{0000-0001-8505-1269}, Michael Buehlmann$^{2}$\orcidlink{0000-0002-8469-4534}, Nicholas Frontiere$^{2}$\orcidlink{0009-0005-8598-4292}, Andrew P. Hearin$^{1}$\orcidlink{0000-0003-2219-6852}, Katrin Heitmann$^{1}$\orcidlink{0000-0003-1468-8232}, Sara Ortega-Martinez$^{1,5}$\orcidlink{0000-0002-9350-8537}, Alan Pearl$^{1}$\orcidlink{0000-0001-9820-9619}, Esteban Rangel$^{2}$\orcidlink{0000-0003-2330-9820}, Silvio Rizzi$^{3}$\orcidlink{0000-0002-3804-2471}, Thomas Uram$^{3}$\orcidlink{0000-0002-5631-0142}, Enia Xhakaj$^{1}$\orcidlink{0009-0006-0366-6037}}

\affiliation{$^1$High Energy Physics Division, Argonne National Laboratory, 9700 South Cass Avenue, Lemont, IL 60439, USA}
\affiliation{$^2$Computational Science Division, Argonne National Laboratory, 9700 South Cass Avenue, Lemont, IL 60439, USA}
\affiliation{$^3$Argonne Leadership Computing Facility, Argonne National Laboratory, 9700 South Cass Avenue, Lemont, IL 60439, USA}
\affiliation{$^4$Institute of Space Sciences (ICE, CSIC), Campus UAB, Carrer de Can Magrans, s/n, 08193 Barcelona, Spain}
\affiliation{$^5$Donostia International Physics Centre, Paseo Manuel de Lardizabal 4, 20018 Donostia-San Sebastian, Spain}

\thanks{$^{\star}$E-mail: gbeltzmohrmann@anl.gov}

\shortauthors{Beltz-Mohrmann et al.}

%%%%%%%%%%%%%%%%%%%%%%%%%%%%%%%%%%%%%%%%%%%%%%%%%%
\begin{abstract}
In this paper, we present the Discovery simulations: a new pair of high-resolution N-body simulations motivated by the DESI Y1 BAO cosmological constraints on dark energy. The Discovery simulations were run with identical initial conditions, and differ only in their cosmological parameters. The first simulation is based on a flat \lcdm cosmology, while the second is based on a \wowacdm cosmology, with particular parameter values chosen based on the DESI analysis which includes constraints from BAO with CMB priors. Both simulations evolve $6720^3$ particles in a box with a side length of $L_\mathrm{box} = 1.5$ Gpc, leading to a mass resolution of $\sim4 \times 10^8$ \Msun in each simulation. In this work we demonstrate the impact of the \wowacdm cosmology on the matter power spectrum, halo mass function, and halo mass accretion rate. We also populate halos with galaxies using a novel forward model for in-situ star formation, and examine the way in which changes to cosmology manifest as changes in star formation history. The Discovery simulations provide a testbed for alternative cosmological probes that may offer additional constraining power beyond BAO, such as higher-order summary statistics and observables in the nonlinear regime. Halo catalogs from the Discovery simulations are publicly available and can be downloaded from the HACC Simulation Data Portal.
\end{abstract}
%%%%%%%%%%%%%%%%%%%%%%%%%%%%%%%%%%%%%%%%%%%%%%%%%%
\maketitle

\vspace{1cm}

\twocolumngrid

\section{Introduction}
\label{section:intro}

Over two decades ago, observational evidence established that the cosmological expansion of the universe is accelerating \citep{riess_etal98,1999ApJ...517..565P,spergel_etal03_wmap1}. This discovery has since motivated a world-wide campaign of large, specially designed astronomical surveys aiming to uncover the physical nature of the dark energy driving the cosmic acceleration. In the mid-2000s, the Dark Energy Task Force (DETF) was formed to help plan and forecast the effort that was to come, and the DETF report organized the progression of future dark energy experiments into a sequence of stages, I-IV, defined by constraining power on dark energy models \citep{detf_report_2006}. One of the principal conclusions of the DETF report was that no single technique can definitively answer the key outstanding questions about dark energy, which will require a coordinated observational and theoretical program over a decades-long effort.

With the arrival of data from the Dark Energy Spectroscopic Instrument \citep[DESI,][]{desi_experiment_design1_2016,desi_instrumentation_overview_2022}, we are now in the era of Stage IV dark energy experiments. The analysis of the Baryonic Acoustic Oscillation (BAO) signature measured from the first year of DESI observations \citep[DESI Y1,][]{desi_y1_paper6_bao_cosmology} suggests that the cosmic acceleration might not be driven by a simple cosmological constant, $\Lambda,$ but instead by a cosmic fluid with an equation of state parameter, $w$, that evolves in time.
This preference for an evolving dark energy model, \wowacdm, has also been found by the Dark Energy Survey \citep[DES,][]{DES2025}.
If confirmed, these findings would have profound implications for cosmology, as these two scenarios for accounting for the cosmic acceleration have dramatically different physical interpretations. However, the initial results from DESI Y1 are inconclusive on this question. Future analyses of DESI data will improve both statistical precision and treatment of systematic errors, and so it is entirely possible that BAO measurements of DESI Y3 will favor $\Lambda$CDM over evolving dark energy; it is also possible that the BAO signal in DESI is simply unable to conclusively discriminate between these two scenarios, and that alternative cosmological probes are needed to shed further light on the physics of dark energy.

Cosmology experiments have come to rely in an essential way upon numerical simulations to deliver their constraints, and use-cases are quite diverse. Simulations are ubiquitously used to mitigate systematic effects in observations, e.g., correcting for fiber collisions in two-point clustering \citep[e.g.,][]{guo_etal12_fiber_collisions,bianchi_etal18_fiber_collisions_desi}, and accounting for telescope point spread functions in measurements of shear \citep{liaudat_etal23_psf_review}. Analyses of observables in the nonlinear regime require suites of simulations for emulator-based predictions \citep{heitmann_etal06_cosmic_calibration,knabenhans_etal23_euclidemulator,storey_fisher_24_aemulus6}. Even predictions for observables in the quasi-linear regime such as BAO require expansive simulation campaigns to achieve a precision that meets the quality of Stage IV cosmology data \citep{padmanabhan_etal12_reconstruction,paillas_etal24_desi_bao_reconstruction}.

Progress in the quality of cosmological N-body simulations has been especially rapid over the last decade. The simplest marker of this progress is the mass resolution that characterizes the simulations used by Stage IV relative to Stage III experiments. Within DESI, the AbacusSummit simulations \citep{Maksimova_etal21_abacus_summit} span a wide range of cosmological parameters, and the primary suite of boxes has a box size of $\Lbox\approx2$\hgpc and a particle mass of $\mptcl\approx2 \times 10^9$\hMsun. Within the Vera C.\ Rubin Observatory Legacy Survey of Space and Time \citep[LSST\footnote{\url{https://rubinobservatory.org}},][]{lsst_science_ivezic_2019}, the N-body simulation Outer Rim \citep{heitmann_etal19_outer_rim} with $\Lbox=3$\hgpc and $\mptcl=1.85 \times 10^9$\hMsun was used in the production of the DC2 synthetic galaxy catalog \citep{lsst_dc2}. The Outer Rim was also used by the OpenUniverse collaboration to produce the Roman-Rubin image simulation \citep{openuniverse_roman_rubin_sim} in collaborative work between LSST and the Nancy Grace Roman Space Telescope\footnote{\url{https://roman.gsfc.nasa.gov}}. Within the Euclid collaboration\footnote{\url{https://www.cosmos.esa.int/web/euclid}} \citep{euclid_definition_2011}, the flagship simulation \citep{euclid_flagship_mock} has $\Lbox=3.6$\hgpc and $\mptcl=1 \times 10^9$\hMsun. 

By contrast, consider the simulation suites used to support Stage III missions. For the case of the Dark Energy Survey \citep[DES\footnote{\url{https://www.darkenergysurvey.org/}},][]{des_y1_cosmology_3x2}, the Aemulus simulations \citep{derose_etal19_aemulus1} were among the principal boxes used to derive cosmological constraints, which have $\Lbox\approx1$\hgpc and $\mptcl\approx3 \times 10^{10}$\hMsun. For the case of the Subaru Hyper Suprime-Cam Survey \citep[HSC\footnote{\url{https://hsc.mtk.nao.ac.jp/ssp/}},][]{aihara_etal18_hsc}, the DarkQuest simulations \citep{nishimichi_etal19} were used, which have $\Lbox\approx1-2$\hgpc and $\mptcl\approx1 \times 10^{10}$\hMsun. 

Based on the above examples, we arrive at a straightforward rule of thumb: contemporary dark energy experiments all use simulations with $\Lbox\approx1-2$\hgpc, but with an order-of-magnitude difference in resolution, $\mptcl=1-2 \times 10^{10}$\hMsun for Stage III, and $\mptcl=1-2 \times 10^9$\hMsun for Stage IV.

In addition to progress in mass resolution, the past decade of cosmological N-body simulations has also seen major improvements in the quality of the higher-level data products. In Stage III simulation suites such as Aemulus, Dark Quest, and Mira-Titan \citep{heitmann_etal_mira_titan}, the simulated data products were typically limited to catalogs of host dark matter halos, without well-resolved substructure. This situation has changed in the Stage IV era. Gpc-scale simulations such as Uchuu \citep[][$\mptcl=3.3 \times 10^8$\hMsun]{ishiyama_etal21_uchuu} and Farpoint \citep[][$\mptcl=4.6 \times 10^7$\hMsun]{frontiere_etal22_farpoint} have not only improved mass resolution by an order of magnitude relative to typical Stage IV suites, but the associated data products include well-resolved substructure and merger trees. This improvement in higher-level data products is not merely attributable to improved computing power, but also to the sophistication of algorithms and code bases; specialized techniques typically need to be developed in order to derive substructure merger trees in Gpc-scale simulations \citep[e.g.,][]{rangel_etal20_core_trees,bose_etal22_abacus_merger_trees,tokuue_etal24_mpi_rockstar}.

Progress in the quality of simulated data products has helped fuel improvements in the realism of models of the galaxy--halo connection in recent years. The Halo Occupation Distribution \citep[HOD,][]{berlind_weinberg_hod_2002} and Conditional Luminosity Function \citep[CLF,][]{yang_etal03_clf} were first developed in the early 2000s, before substructure catalogs had become standard, and so for years HOD and CLF implementations never utilized more than catalogs of host halos defined by their mass. As simulated data products improved, HOD/CLF models began to include additional dependence upon halo assembly history and other properties \citep[referred to as {\it galaxy assembly bias}, see, e.g.,][]{croton_etal07_assembias,zentner_hearin_vdb_14_assembias,hearin_etal16_decorated_hod}, and by now it has become standard to include such effects into halo occupation models \citep[e.g.,][]{wibking_etal19,yuan_etal22_abacushod,rocher_etal23}. 

Models of the galaxy--halo connection that directly use substructure merger trees have also improved and proliferated in recent years. The subhalo abundance matching technique \citep[SHAM,][]{conroy_etal06_sham} maps synthetic galaxies onto subhalos, and the most successful implementations are based on historical (sub)halo properties derived from the merger trees \citep[e.g.,][]{reddick_etal13,chaves_montero_etal16_vrelax}. SHAM-type models are now widely used within DESI \citep[e.g.,][]{prada_etal23_desi_onepct_uchuu,Gao2023,ereza_etal24_uchuu_glam,Yu2024}, and have become a science driver of some cosmological simulation suites \citep{angulo_etal21_bacco,contreras_etal21_shame}. 

Nearly all contemporary semi-analytic models of galaxy formation \citep[SAMs,][]{benson_galaxy_2012,overzier_etal13,henriques_etal17,gonzalez_perez_etal18_galform17,lagos_etal18,delucia_etal24_gaea_sam} are based on substructure merger trees, and SAMs have also greatly benefited from improvements to simulated data products in recent years \citep[see][for a review]{somerville_physical_2015}. Recently, models dubbed semi-empirical forward models \citep{wechsler_tinker18} are becoming more widely used to generate synthetic data for cosmological analyses \citep{behroozi_etal19_umachine,aung_etal23_uchuu_umachine,openuniverse_roman_rubin_sim}; these models bridge the gap between SAMs and halo occupation models by leveraging substructure merger tree information, but adopting computationally efficient prescriptions for generating synthetic galaxies. 

Hydrodynamical simulations provide another method for modeling the formation and evolution of galaxies. These simulations include parameters that capture baryonic physics via phenomenological subgrid models. However, hydrodynamical codes with different implementations of subgrid models and choices of modeling parameters can exhibit substantial disagreement in results. Additionally, the choices of modeling parameters must be adjusted each time the resolution of the simulation is changed, in order to continue to obtain accurate agreement with observations used for calibration. The high computational cost of these simulations presents an additional difficulty, limiting our ability to use them for cosmological inference. For these reasons, the generation of synthetic sky catalogs for cosmological surveys still relies mostly on gravity-only simulations and flexible modeling of the galaxy-halo connection in post-processing.

Motivated by the recent DESI results, as well as recent developments in cosmological simulations, in this paper we present the Discovery simulations: two high-resolution, Gpc-scale cosmological N-body simulations that we have designed to provide a testbed for studies of dark energy that follow up on recent DESI results. The Discovery simulations are a pair of boxes with identical initial conditions; the only difference between the two simulations is cosmology. The first box uses a $\Lambda$CDM cosmology, while the second box contains a dark energy equation of state $w$ which evolves in time. The particular values of the cosmological parameters in each simulation were chosen to closely mimic the best-fitting cosmologies in the DESI Y1 BAO analyses (see \S\ref{sec:sims} for details).
Data products for the Discovery simulations include halo catalogs as well as substructure merger trees.

Our goal for this work is to present the Discovery simulations, and to illustrate the kind of dark energy sensitivity analyses that they enable.
We aim for these simulations to form the basis of future synthetic galaxy catalogs, as well as studies of the ways in which two-point and higher-order summary statistics in the nonlinear regime depend upon dark energy.

This paper is organized as follows. In \S\ref{sec:sims}, we describe the Discovery simulations and associated data products. In \S\ref{sec:results}, we present a comparison of the two simulations, including matter power spectra in \S\ref{subsec:pk}, halo mass functions in \S\ref{subsec:mf}, and halo mass assembly histories in \S\ref{subsec:mah}; in \S\ref{subsec:sfh} we use a novel forward model of the galaxy-halo connection to estimate how the difference in cosmology between the two simulations manifests in central galaxy star formation history. We conclude in \S\ref{sec:conclusion} with an outline of future work, and a summary of our principal results.
%%%%%%%%%%%%%%%%%%%%%%%%%%%%%%%%%%%%%%%%%%%%%%%%

\section{Simulations}
\label{sec:sims}

The Discovery simulations presented in this paper were carried out with the Hardware/Hybrid Accelerated Cosmology Code (HACC), described in~\cite{2016NewA...42...49H}. HACC has been optimized to run at scale on all currently available high-performance computing architecture. As it is a common approach in cosmological N-body solvers, the HACC solver is split into a short-range and a long-range part with a smooth hand-over between scales. The HACC long-range solver is based on a particle-mesh method using large FFTs, and is not changed for different high-performance computing systems, while the short-range solver is optimized for different computational architectures \citep{habib2013hacc}. For accelerator-based systems, e.g. GPU-based systems, HACC uses a direct particle-particle solver on short distances, while for non-accelerated systems, tree structures are built to obtain excellent performance on small scales. HACC is fully equipped to carry out major analysis tasks in situ, and several analysis tools, e.g. the halo finder, utilize GPU-acceleration. 

The Discovery simulations ran on 960 nodes (5760 GPUs) of Aurora\footnote{\url{https://www.alcf.anl.gov/aurora}} ($\sim$ 10\% of the full machine), a GPU-accelerated exascale supercomputer hosted at the Argonne Leadership Computing Facility\footnote{\url{https://www.alcf.anl.gov/}} (ALCF). The run time for each simulation, including the in-situ analysis and I/O, was $\mathcal{O}$(2 days), enabling close-to-real-time investigations of new cosmological results.

\begin{figure}
\includegraphics[width=\columnwidth]{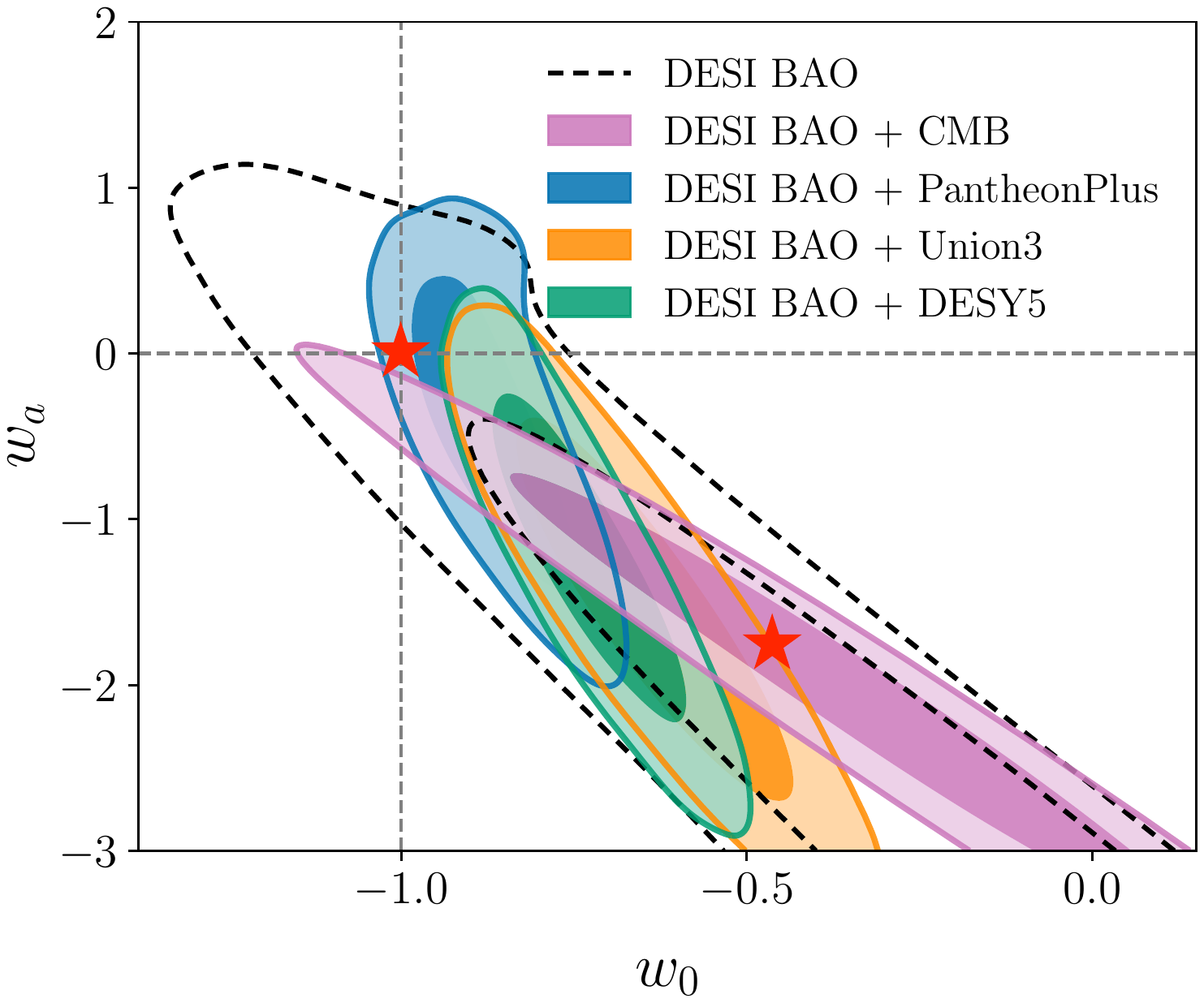}
\caption{{\bf Illustration of the $w_0$ and $w_a$ parameters used in the Discovery simulations (red stars), compared to the DESI Y1 BAO analysis.} The contours (reproduced from Fig. 6 of~\citealt{DESI_2024}) show the 68\% and 95\% marginalised posterior constraints for the flat \wowacdm model from DESI BAO alone (black dashed), DESI + CMB (pink), and DESI + SN Ia, for the PantheonPlus, Union3, and DESY5 SNIa datasets in blue, orange and green respectively.}
\label{fig:desi}
\end{figure}

\subsection{Parameter Choices}

In this section we briefly discuss our choices for the simulation settings and cosmological parameters of the Discovery Simulations. In contrast to hydrodynamical simulations, gravity-only simulations have no free modeling parameters. Settings for, e.g., resolution, simulation volume, and starting redshifts, have been extensively discussed in the literature to achieve sub-percent level accuracy to at least $k\sim 1$Mpc$^{-1}$ and high accuracy beyond \citep[see, e.g.,][]{2010ApJ...715..104H,2016JCAP...04..047S, 2021MNRAS.505.2840E}. Additionally, code comparisons have been performed in multiple studies \citep[e.g.,][]{2008CS&D....1a5003H,2022MNRAS.515.1854G}. A recent summary of these comparisons can be found in \cite{2022LRCA....8....1A}.

The simulation parameters used in the Discovery Simulations have been carefully chosen to meet the accuracy requirements derived in these recent publications. Each simulation covers a volume of (1500Mpc)$^3$ and evolves 6720$^3$ particles with a force resolution of $\sim 2$kpc. We begin the simulations at $z=200$ and use the Zel'dovich approximation to assign initial particle positions and velocities \citep{1970A&A.....5...84Z}. We generate input transfer functions with CAMB \citep{2000ApJ...538..473L}. Our approach for generating initial conditions is described in \cite{2010ApJ...715..104H}, which found that the Zel'dovich approximation and the 2LPT approach yield very similar initial displacements and velocities at high redshift, and converge at $z=0$.

We have chosen cosmological parameters for the Discovery simulations based on the recent DESI Y1 analysis of BAO with CMB priors~\citep{DESI_2024}.
In Fig.~\ref{fig:desi} we show the DESI constraints in the $w_0w_a$ plane for a flat \wowacdm model (reproduced from Fig. 6, left panel in \citealt{DESI_2024}). 
Specifically, we determine our parameters using the results that combine the DESI BAO measurements with constraints from the CMB (pink contours).
The values of $w_0$ and $w_a$ that we use in the Discovery simulations are marked by the red stars. We follow the convention that the evolving dark energy equation of state is parameterized via $w(a) = w_0 + w_a(1 - a)$.

For the \lcdm~simulation, we use the following cosmological parameters:
\begin{eqnarray}
\Omega_\mathrm{m}&=&0.3069,\nonumber\\ 
\Omega_\mathrm{b} h^2 &=&0.0225,\nonumber\\ 
h&=&0.6797,\nonumber\\ 
\sigma_8&=&0.8135,\nonumber\\ 
n_s&=&0.968,\nonumber\\
w_0&=&-1,\nonumber\\
w_a&=&0,
\end{eqnarray}
leading to a mass resolution of $m_{\rm ptcl}=4.38 \times 10^8$\Msun ($2.98 \times 10^8$\hMsun). The choice of $\sigma_8$ normalization corresponds to a primordial power spectrum amplitude $A_s=2.059\times 10^{-9}$ fixed at $k=0.05h$Mpc$^{-1}$.

For the \wowacdm simulation, we use the following parameters:
\begin{eqnarray}
\Omega_\mathrm{m}&=&0.3439,\nonumber\\ 
\Omega_\mathrm{b} h^2 &=&0.0224,\nonumber\\ 
h&=&0.6466,\nonumber\\ 
\sigma_8&=&0.7917,\nonumber\\ 
n_s&=&0.965,\nonumber\\
w_0&=&-0.45,\nonumber\\
w_a&=&-1.79,
\end{eqnarray}
leading to a mass resolution of $m_{\rm ptcl}=4.42 \times 10^8$\Msun ($2.86 \times 10^8$\hMsun). The choice of $\sigma_8$ normalization corresponds to a primordial power spectrum amplitude $A_s=2.003\times 10^{-9}$ fixed at $k=0.05h$Mpc$^{-1}$. For both simulations, we assume a flat cosmology and massless neutrinos, $\omega_\nu=0$.

\begin{figure}
\includegraphics[width=\columnwidth]{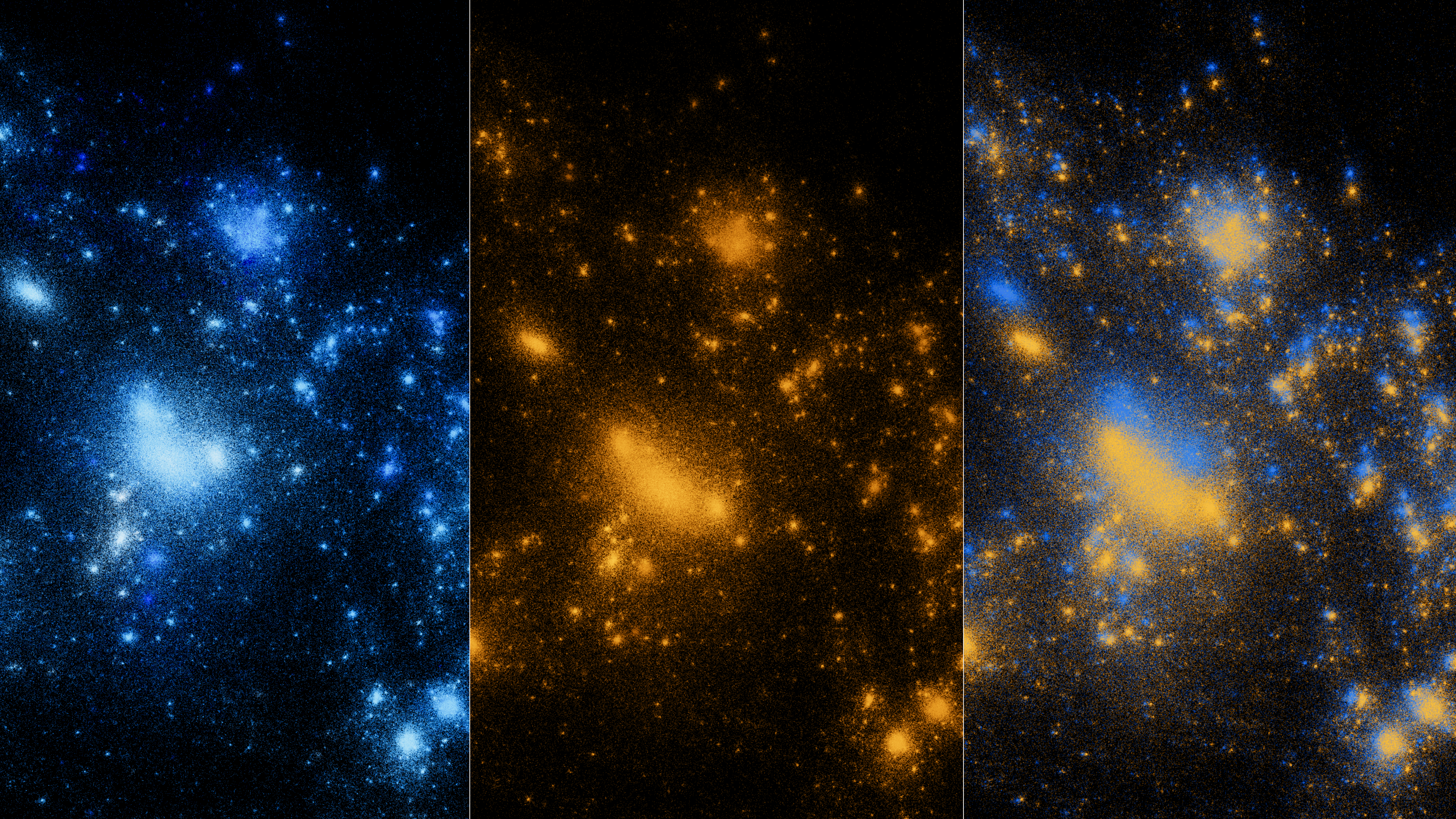}
\caption{{\bf A visual comparison of a small region in the simulations at $z=0$.} Left: \lcdm (blue); Center: \wowacdm (yellow); Right: an overlay of the two simulations. The differences are subtle but still clearly visible at the substructure level, particularly when overlaid. This comparison demonstrates the precision cosmology challenge to obtain cosmological measurements that are sensitive to the smallest changes in structure formation. A short movie showing the evolution of a small subvolume of the simulation is available \href{https://anl.box.com/s/lzulgzw5mc1tsy8bmndm3i9fkttblqqo}{here}.}
\label{fig:vis}
\end{figure}

By combining the large, survey-scale volumes of the simulations with high-resolution in mass, we are able to extract substructure merger trees and build detailed synthetic skies. These simulations therefore provide a unique resource for the cosmology community to further investigate measurements obtained from ongoing and upcoming surveys. 
In Fig.~\ref{fig:vis} we compare a small region of the two Discovery simulations to illustrate the subtle differences in substructure due to the small differences in cosmological parameter choices.
This image presents a visual glimpse of the challenge for precision cosmology, and demonstrates the importance of high resolution simulations for studying differences in structure formation on small scales.

\subsection{Merger Trees and Substructure Tracking}

To construct detailed merger trees from the simulations, we identify friends-of-friends (FOF) halos \citep{1985ApJ...292..371D} at 101 time steps starting at $z\sim 10$, evenly spaced in $\log_{10}a$. We choose a linking length of $b=0.168$ and capture halos with at least 20 particles. The choice for the linking length was guided by previous work on the generation of synthetic sky catalogs for spectroscopic surveys \citep{2011MNRAS.417.1913R,2011ApJ...728..126W}.  For each halo with 80 particles or more, we register 50 particles closest to the center, the halo core. Each particle in the simulation has a unique tag which does not change throughout the simulation. Using this tag, we track each particle that is at any point identified as a core particle throughout the rest of the simulation. This allows us to continue to follow the evolution of halo cores even after infall into other halos and, therefore, to capture information about substructure. \cite{2021ApJ...913..109S} provide an in-depth comparison with a conventional subhalo finding approach and \cite{2024arXiv240700268S} explore the connections of cores and galaxies in hydrodynamical simulations. \cite{2023OJAp....6E..24K} demonstrate how this approach can be used to accurately model the distribution of galaxies in clusters. 

The particle tags, as well as the halo and core information, are the key ingredients for building detailed merger trees that include substructure information. 
This ``core" approach for tracking substructure growth was introduced and is described in detail in~\cite{rangel_etal20_core_trees}.
The two main advantages of the core approach compared to more traditional subhalo finding are (i) computational efficiency (identifying halo cores and tracking them over time while the simulation is evolving adds almost no computational cost to the simulation), and (ii) providing an approach for tracking substructure even if a subhalo gets disrupted after falling into a host halo.
One downside of not directly identifying subhalos is that some additional subhalo information is lost; however, some of this information can be modeled in post-processing, as shown in~\cite{2021ApJ...913..109S}.

In this work, we use the core merger trees from the two Discovery simulations to investigate the impact of an evolving dark energy on the mass assembly history of halos (see \S~\ref{subsec:mah}), and subsequently how differences in halo mass assembly history manifest themselves as differences in the galaxy population, using a novel model of the galaxy-halo connection (see \S~\ref{subsec:sfh}).

\subsection{Data Release}

As part of this paper we publicly release FOF halo catalogs at three redshifts: $z=1.0, 0.5,$ and 0. For each halo we provide a center position in comoving $h^{-1}$Mpc measured with two different algorithms. The first center position is determined from the gravitational potential minimum of the halo; the second center position provides information about the center of mass of the halo. The results of these two measurements can vary significantly for large unrelaxed halos and can be used to define a relaxation criterion for halos \citep{2018ApJ...859...55C}. We also provide the halo comoving peculiar velocities, measured in km/s. Finally, we supply the mass of each halo in units of \hMsun, measured by simply counting the number of particles belonging to the FOF halo. The data can be downloaded from the HACC Simulation Data Portal\footnote{\url{https://cosmology.alcf.anl.gov/}} with a Globus account.

\section{Results}
\label{sec:results}
One of the main goals of this work is to illustrate the comparisons between \lcdm and \wowacdm that are enabled by these two simulations.
In this section, we examine differences in the matter power spectra at three redshifts ($z=1.0,0.5,$ and 0) between the two cosmologies.
We also investigate how evolving dark energy affects the halo mass functions and halo mass assembly histories between $z=5$ and $z=0$.
Finally, we examine how changes in halo mass accretion rates propagate to changes in galaxy star formation histories in a \wowacdm universe\footnote{Because the two Discovery simulations use different values of $h$, all comparisons are shown in units of, e.g., \Msun, Mpc, etc. \citep{damnyoulittleh}.}.
It should be noted that because the Discovery simulations contain differences in all of their cosmological parameters, we are not isolating the effect of evolving dark energy, but rather examining the differences in these two simulations based on their overall cosmologies.

\subsection{Matter Power Spectra}
\label{subsec:pk}

The nonlinear matter power spectrum and its evolution over time hold valuable cosmological information. HACC allows for a fast evaluation of the power spectrum on-the-fly, and as part of the simulation runs we have evaluated the power spectrum at many time steps. The power spectrum is not directly measured from the particles, but rather by first generating a density field on a grid using a cloud-in-cell assignment (appropriately normalized by the volume of the simulation) and applying a discrete FFT to obtain $\delta({\bf k})$.  From this we determine the three-dimensional $P({\bf k})=|\delta({\bf k})|^2$ which is then binned in amplitudes and divided by the mode multiplicity in each bin to obtain the one-dimensional matter power spectrum $P(k)$. The smoothing from the CIC assignment is compensated for by dividing the power spectrum by the window function that corresponds to the CIC assignment scheme.

%%%%%%%%%%
\begin{figure}
\includegraphics[width=8.5cm]{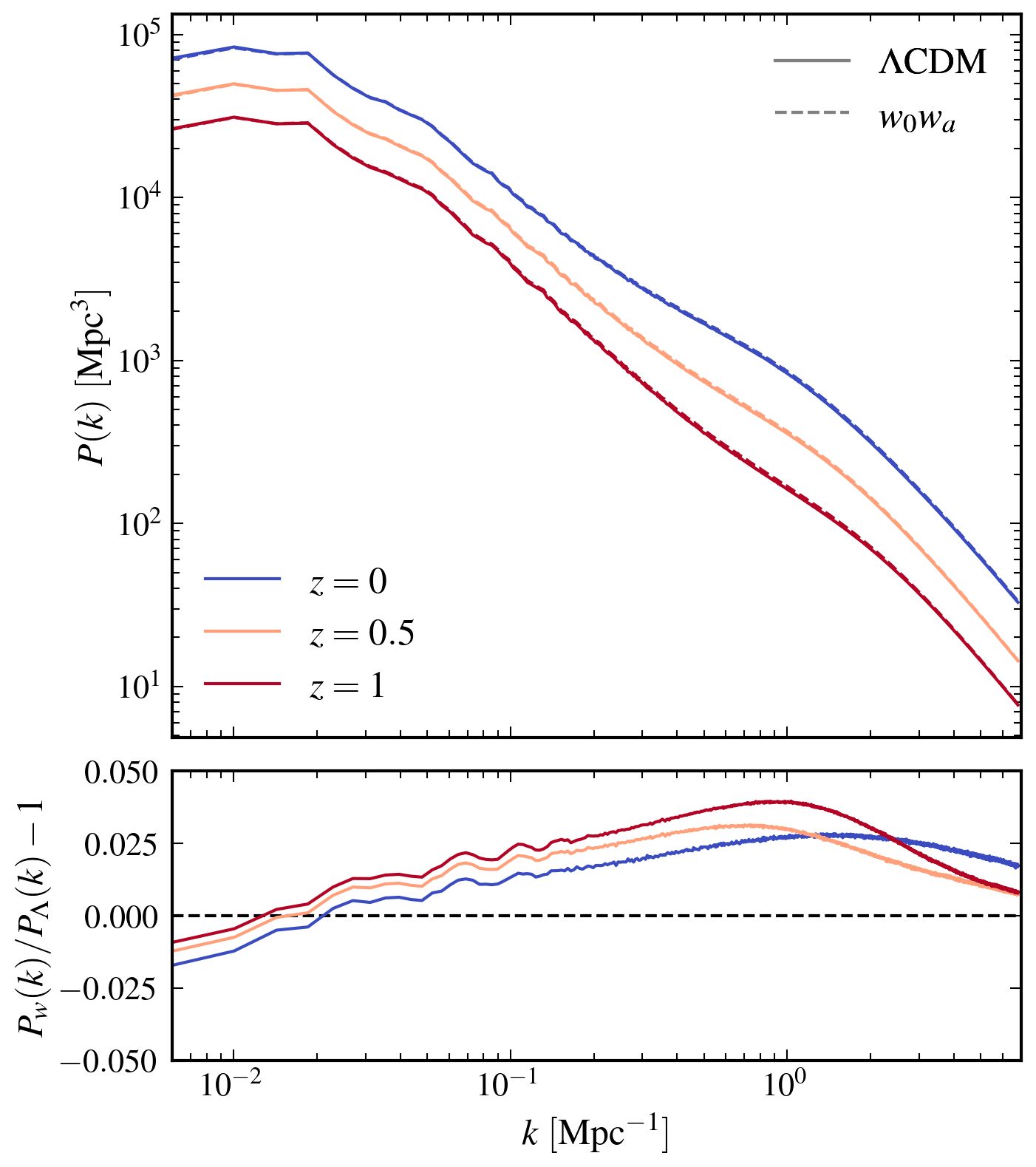}
\caption{{\bf Influence of \wowacdm cosmology on the matter power spectrum.} \textit{Top:} Matter power spectra for the \lcdm simulation (solid) and the \wowacdm simulation (dashed) at $z=0$ (blue), $z=0.5$ (pink), and $z=1$ (red). \textit{Bottom:} Fractional difference in the matter power spectra between the two simulations at each redshift.}
\label{fig:pk}
\end{figure}
%%%%%%%%%%

In Fig.~\ref{fig:pk} we demonstrate the impact of changing cosmology on the matter power spectrum at $z=0$ (blue), $z=0.5$ (pink), and $z=1$ (red).
The bottom panel shows the fractional difference between the matter power spectra in the \wowacdm and \lcdm simulations at each redshift.
Overall, changes in the matter power spectrum are at the level of a few percent.
On the largest scales ($k < 0.02$Mpc$^{-1}$), \wowacdm exhibits a suppression of the power spectrum compared to \lcdm, with the largest deviations appearing at $z=0$ ($\sim$2\%).
Similarly, on the smallest scales ($k > 2$Mpc$^{-1}$),  \wowacdm exhibits an enhancement in the power spectrum at the level of 1-2\%, with the largest differences appearing at $z=0$.
On intermediate scales ($0.02 < k < 2$Mpc$^{-1}$), \wowacdm also leads to an enhancement in the matter power spectrum.
On these scales, the largest deviations from \lcdm appear at $z=1$, with a deviation of $\sim$4\% at $k = 1$Mpc$^{-1}$.

Once again, these differences in the matter power spectrum cannot be exclusively attributed to the influence of evolving dark energy, as the Discovery simulations also contain differences in other cosmological parameters (e.g. $\Omega_m$ is slightly larger and both $\sigma_8$ and $H_0$ are slightly smaller in the \wowacdm simulation compared to the \lcdm simulation.)

\subsection{Halo Mass Functions}
\label{subsec:mf}
\begin{figure}
\includegraphics[width=8.5cm]{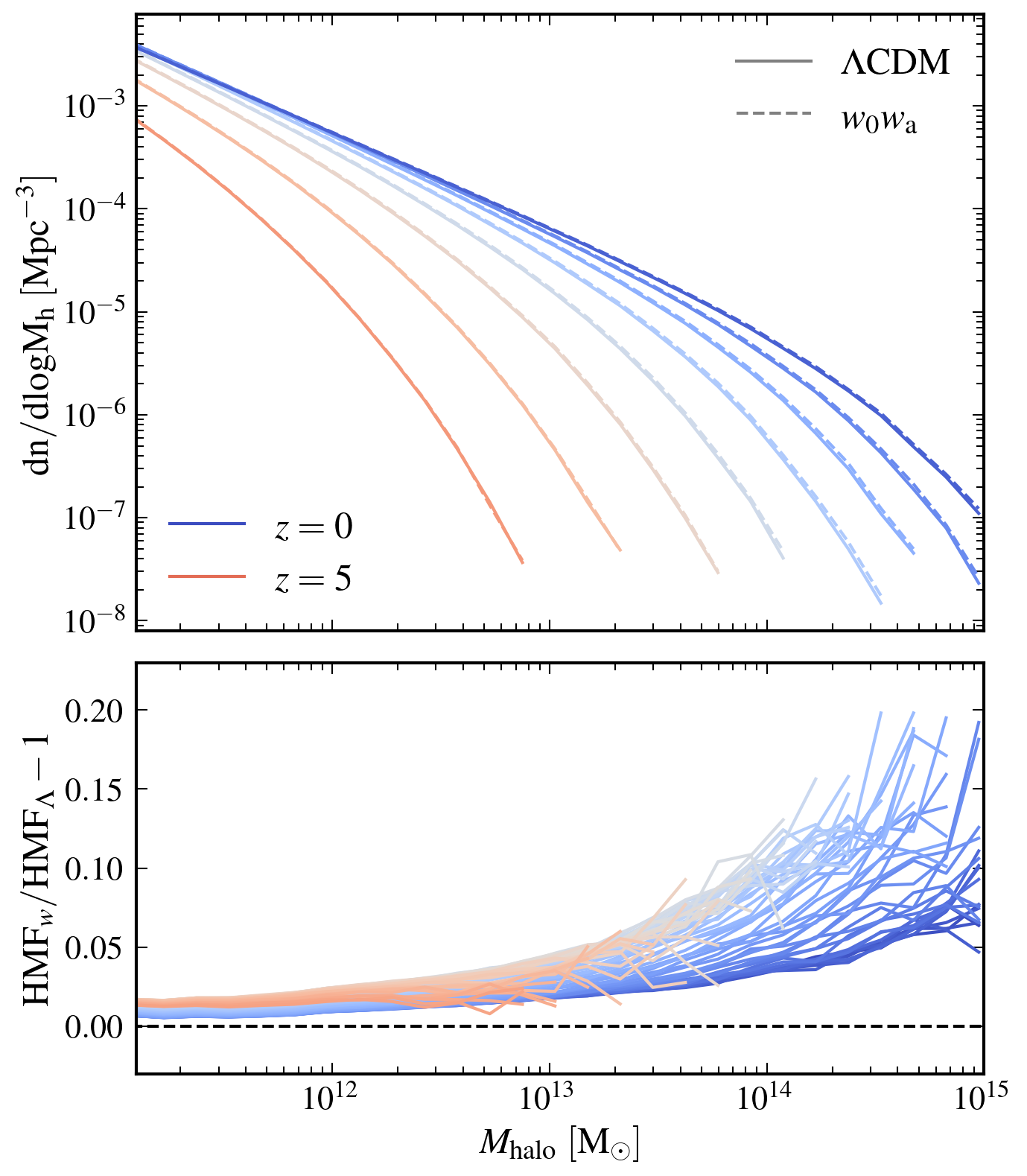}
\caption{{\bf Influence of \wowacdm cosmology on the halo mass function at different redshifts.} \textit{Top:} Halo mass functions in the \lcdm simulation (solid lines) and in the \wowacdm simulation (dashed lines) at several redshifts between $z=5$ (red) and $z=0$ blue. \textit{Bottom:} The fractional difference between the halo mass function in the \wowacdm simulation compared to the \lcdm simulation for a number of redshifts between $z=5$ (red) and $z=0$ (blue). Only mass bins that contain 50 or more halos are shown.}
\label{fig:hmf}
\end{figure}

As detailed in \S\ref{sec:sims}, HACC is equipped with an in situ halo finder, and we have generated FOF halo catalogs with a linking length of $b=0.168$ at 101 time snapshots. In the following section we examine the effect of changing cosmology on the halo mass function measured from a subset of the snapshots. 
In the top panel of Fig.~\ref{fig:hmf} we show the halo mass function in the \lcdm simulation (solid) and in the \wowacdm simulation (dashed) at several redshifts between $z=5$ (red) and $z=0$ (blue).
In the bottom panel we show the fractional difference between the halo mass function in the \wowacdm and \lcdm simulations at each snapshot between $z=5$ and $z=0$.
Only mass bins that contain 50 or more halos are shown.

%%%%%%%%%%

At all redshifts and for all masses the halo mass function is suppressed in the \lcdm simulation compared to the \wowacdm simulation.
For low mass halos ($M_\mathrm{halo} < 10^{12}$ \Msun), the difference in the halo mass function is $\sim$1\% at $z=0$ and $\sim$2\% at $z=5$.
This discrepancy widens as we examine larger halo masses.
For halos with $M_\mathrm{halo} > 10^{14}$ \Msun, we see discrepancies in the mass function anywhere between 5-20\% depending on redshift, although these differences are exacerbated by the rarity of high mass halos.
Nevertheless, it is clear that the \wowacdm cosmology results in an overall shift in the halo mass function toward higher masses compared to the \lcdm cosmology.
Once again, these differences are not purely due to the effects of evolving dark energy, but rather overall changes in the two cosmological models.
For the case of the halo mass function in particular, the shift is likely driven in part by the increase in $\Omega_m$ when switching from \lcdm to \wowacdm.

%%%%%%%%%%
\begin{figure}
\includegraphics[width=8.5cm]{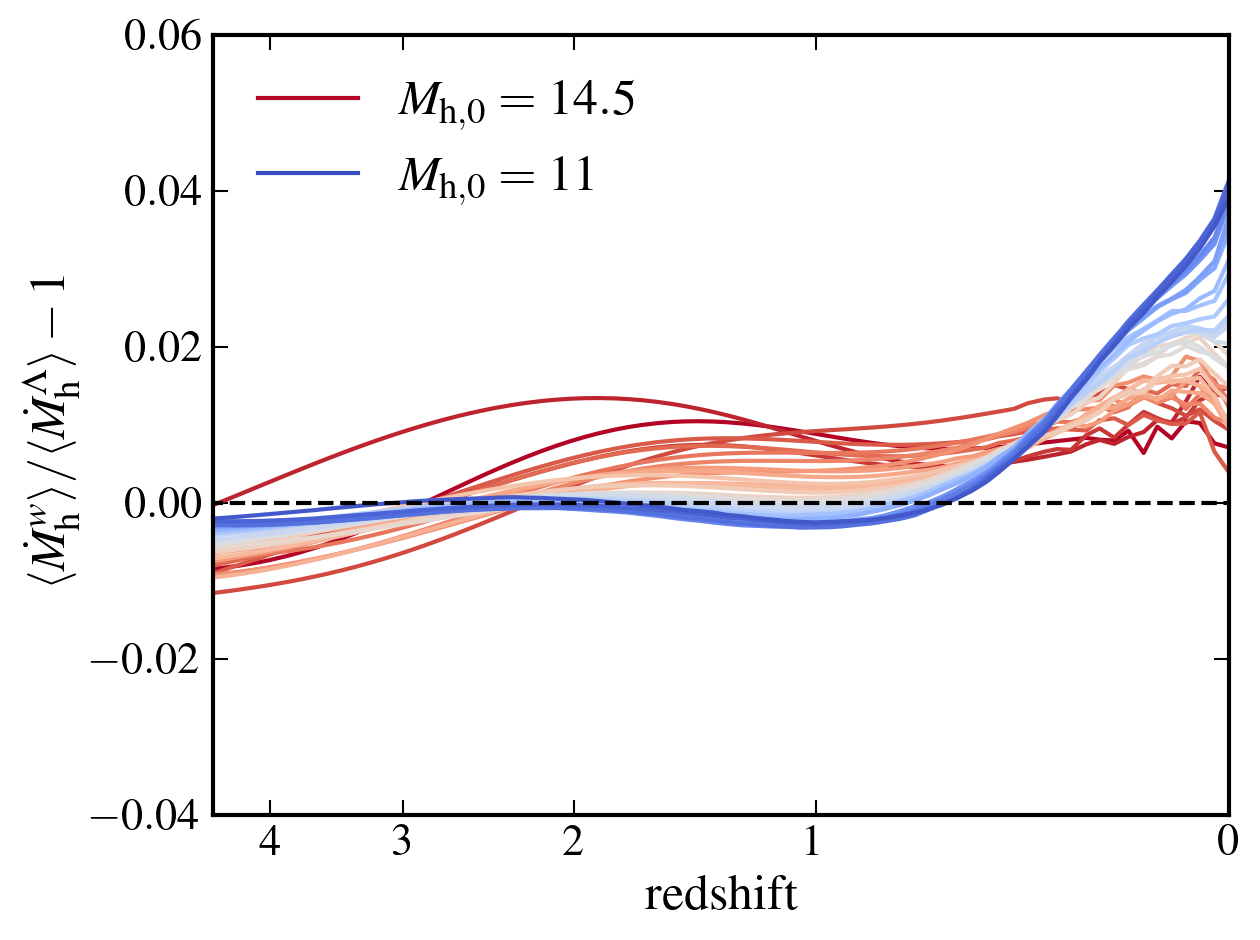}
\caption{{\bf Influence of \wowacdm cosmology on halo accretion rate.} The fractional difference between the average halo mass accretion rate in the \wowacdm simulation compared to the \lcdm simulation is plotted as a function of redshift in several halo mass bins between $M_h(z=0) = 11$ \Msun (blue) and $M_h(z=0) = 14.5$ \Msun (red).}
\label{fig:mah}
\end{figure}
%%%%%%%%%%

\subsection{Mass Assembly of Halos}
\label{subsec:mah}
In addition to examining the impact of cosmology on the halo mass function, we also investigate the influence of \wowacdm on halo mass assembly rates.
Many properties of halos are closely connected to their assembly histories, including concentration \citep{Zhao2003, Wang2020}, substructure abundance \citep{Gao2004, Wechsler2006, Giocoli2010}, the outer density profile \citep{Diemer2014}, and ellipsoidal shape \citep{Chen2020, Lau2021}.
Additionally, there is a strong relationship between the assembly history of a halo and the larger-scale environment in which it evolves, a phenomenon known as halo assembly bias \citep[e.g.,][]{Gao2005, Mao2018, Mansfield2020}.
Moreover, many models of the galaxy-halo connection utilize halo assembly histories for predicting galaxy star formation rates (see Section~\ref{subsec:sfh}).

In Fig.~\ref{fig:mah} we show the fractional difference between the average halo mass accretion rate in the \wowacdm and \lcdm simulations as a function of redshift in several halo mass bins between $M_h(z=0) = 11$ \Msun (blue) and $M_h(z=0) = 14.5$ \Msun (red).
At low redshift ($z < 0.5$), \wowacdm halos exhibit a slightly faster mean mass accretion rate than \lcdm halos, with $10^{11}$ \Msun halos experiencing a $\sim$4\% higher accretion rate at $z=0$ in the \wowacdm simulation. 
Higher mass halos, while still experiencing a faster accretion rate at $z=0$, exhibit smaller deviations from \lcdm (1-2\%).
These deviations from \lcdm at low redshift are to be expected given the advent of the dark energy dominated regime at $z < 1$.
At redshifts greater than $z=1$ the difference between the mean halo mass accretion rates of low mass halos in the two simulations is very small.

For high mass halos, the discrepancy overall in the mean mass accretion rates of the two cosmologies is larger, although the exact difference varies depending on redshift and halo mass.
At intermediate redshifts ($1 < z < 2$) mass accretion rates are $\sim$1\% faster for high mass halos in the \wowacdm cosmology.
At higher redshifts ($z > 3$), halos in the \wowacdm simulation exhibit a slightly slower mass accretion rate than those in the \lcdm simulation.
This effect is more significant for high mass halos, with $10^{14.5}$\Msun halos in the \wowacdm cosmology experiencing a $\sim$1\% lower mass accretion rate at $z=4$ than the same halos in the \lcdm cosmology.
While we expect evolving dark energy to impact the growth of structure primarily at low redshift, differences in $\Omega_\mathrm{m}$ and $\sigma_8$ could play a role in the differences in halo mass assembly rates measured at high redshift.

\subsection{Star Formation Histories of Galaxies}
\label{subsec:sfh}
Having examined the way in which changing cosmology influences the mass assembly history of halos, we consider next the way in which differences in halo mass assembly histories manifest in differences in galaxy star formation histories.
Many theoretical models of galaxy formation rely on mass assembly histories of halos for predicting galaxy star formation rates \citep[e.g.,][]{Kauffmann1993, Somerville2008, Watson2015, Behroozi2019}.

In this work, we investigate the way in which evolving dark energy impacts galaxy formation by applying a parameterized forward model of galaxy formation to both simulations.
The model, Diffstarpop\footnote{\url{https://github.com/ArgonneCPAC/diffstarpop}} (Alarcon et al. in preparation), uses a number of free parameters combined with simulated halo mass accretion rates to generate statistical realizations of galaxy populations.
Diffstarpop is a generative version of Diffstar\footnote{\url{https://github.com/ArgonneCPAC/diffstar}} \citep{Alarcon2023}, a fully parametric physical model for fitting the assembly history of galaxies.
The Diffstar model was developed and validated by comparing to existing models of galaxy formation, including UniverseMachine \citep{behroozi_etal19_umachine} and IllustrisTNG \citep{Marinacci2018,Naiman2018,Nelson2018,Pillepich2018a,Pillepich2018b,Springel2018}.
In ongoing work, a preliminary calibration of the Diffstarpop model parameters have been tuned to reproduce the galaxy population in UniverseMachine.

We apply this preliminary calibration of Diffstarpop to the Discovery simulations to populate both boxes with galaxies.
Prior to applying Diffstarpop, we fit the mass assembly histories of the halos in each simulation with Diffmah\footnote{\url{https://github.com/ArgonneCPAC/diffmah}} \citep{Hearin2021}, a differentiable empirical model for smoothly approximating the growth of halos.
This preprocessing step allows us to summarize the assembly history of each individual halo with just four parameters describing the growth of the halo over time.
Subsequently, we use the smooth Diffmah fit for each halo to map parameters onto each halo describing its star formation history.
In this way we are able populate the halos in each simulation with galaxies which have star formation histories that are directly related to their halo's mass assembly history.

%%%%%%%%%%
\begin{figure}
\includegraphics[width=8.5cm]{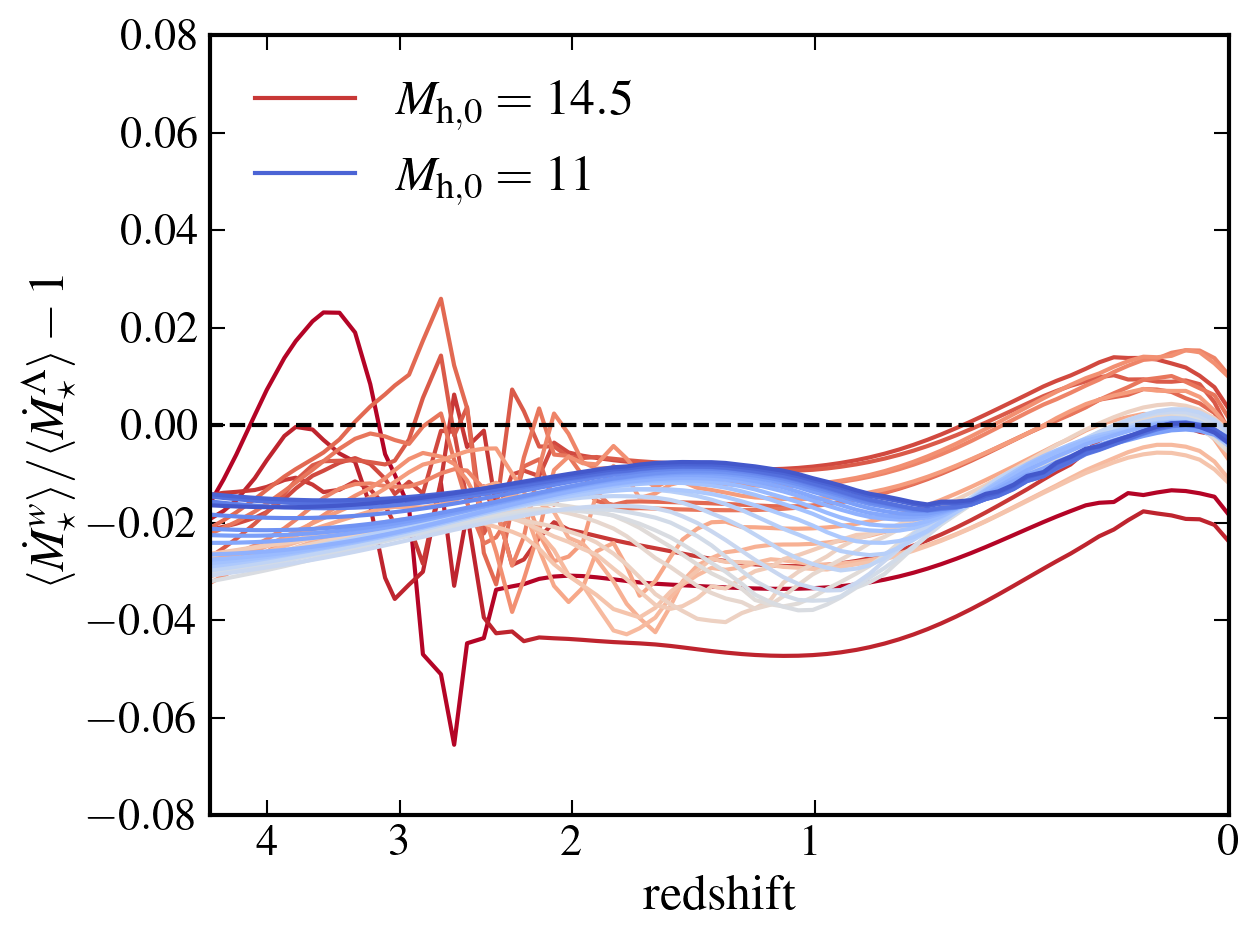}
\caption{{\bf Influence of \wowacdm cosmology on star formation rate.} The fractional difference between the average star formation rate in the \wowacdm simulation compared to the \lcdm simulation is plotted as a function of redshift in several halo mass bins between $M_h(z=0) = 11$ \Msun (blue) and $M_h(z=0) = 14.5$ \Msun (red).}
\label{fig:sfh}
\end{figure}
%%%%%%%%%%

We use this novel galaxy-halo connection model to populate each of the Discovery simulations with galaxies.
In Fig.~\ref{fig:sfh} we show the fractional difference in average star formation rate for central galaxies in the \wowacdm and \lcdm simulations as a function of redshift in several halo mass bins between $M_h(z=0) = 11$ \Msun (blue) and $M_h(z=0) = 14.5$ \Msun (red).
In this figure, the same Diffstarpop model is applied to both simulations; in other words, differences in the mean star formation rates are due to differences in cosmology and halo mass assembly, and not due to differences in star formation parameters.

Among low mass halos, the largest differences in star formation rates between the two cosmologies appears at high redshift ($z>1$), where the star formation rate of \wowacdm galaxies is $\sim 2$\% lower than the star formation rate of \lcdm galaxies.
Among intermediate mass halos, we also see $\sim 3$\% differences in star formation rates around $z=1$.
For high mass halos, we see differences of up to 4\% in the galaxy star formation rates. 
However, it is noteworthy that the rarity of very high mass halos leads to a substantial amount of scatter in the comparisons of star formation rates.

The trends shown in Fig.~\ref{fig:sfh} are driven by a combination of three factors: the differences in baryon fraction between the two cosmologies, the differences in halo mass assembly rates between the two cosmologies, and the particular star formation model that was applied to the Discovery simulations in this work.
The Diffstar model assumes that $\mathrm{d}M_\mathrm{g}/\mathrm{d}t \propto f_\mathrm{b} \times \mathrm{d}M_\mathrm{h}/\mathrm{d}t$, where $\mathrm{d}M_\mathrm{g}/\mathrm{d}t$ is the accretion rate of baryonic material that is available for star formation, $\mathrm{d}M_\mathrm{h}/\mathrm{d}t$ is the growth rate of total halo mass, and $f_\mathrm{b}$ is the baryon fraction in the simulation.
The baryon fraction in the Discovery simulations is 0.1587 in the \lcdm cosmology and 0.1558 in the \wowacdm cosmology, a difference of $\sim 2$\%.
This difference in baryon fraction explains the overall $\sim 2$\% offset in star formation rates between the two simulations at $z>1$, with the \wowacdm cosmology exhibiting a slower star formation rate due to its lower baryon fraction.
Had the Discovery simulations been run with the exact same baryon fraction, we would likely see no substantial offset in star formation rates at high redshift, and discrepancies at lower redshift would become more apparent.

Additionally, it is important to note that certain parameter choices in the star formation model can lead to star formation histories which are more or less sensitive to changes in halo mass accretion rates over time.
For example, one of the parameters of our star formation model is $\tau_\mathrm{cons}$, the timescale over which an accreted parcel of gas
will be gradually transformed into stellar mass.
The star formation rate of any individual galaxy, $dM_\star (t)/dt$, receives a contribution from all of the previously accreted gas, $M_g(t')$, for all times $t - \tau_\mathrm{cons} \leq t' \leq t$.
The larger the value of $\tau_\mathrm{cons}$, the longer this gas consumption timescale.
Evidence for long gas consumption timescales can be found in detailed analyses of high-resolution hydrodynamical simulations \citep[e.g.,][]{Semenov2017,Semenov2018}, as well as from observations of the Milky Way and nearby spiral galaxies \citep[e.g.,][]{Kennicutt1989, Kennicutt1998, Bigiel2008, Leroy2008, Leroy2013, DeLosReyes2019, DiazGarcia2020, Kennicutt2021}.
A longer gas consumption timescale means that a galaxy's star formation rate at any given time depends on a longer baseline of the mass assembly history, whereas a shorter gas consumption timescale produces star formation rates that depend only on the very recent mass assembly history.
Thus, a star formation model with a long gas consumption timescale might be more sensitive to changes in cosmology that impact halo mass assembly history, even if those changes are subtle.
Here we have provided a demonstration of how the combination of \wowacdm with a particular star formation model impacts galaxy star formation rates; in future work we will use DiffstarPop to programmatically vary the star formation model to explore the interplay between cosmology and galaxy formation physics.

\section{Conclusion}
\label{sec:conclusion}
In this work, we present the Discovery simulations: two new N-body simulations motivated by the DESI Y1 BAO constraints on evolving dark energy.
One simulation was run with a flat \lcdm cosmology, while the other was run with a \wowacdm cosmology.
The specific cosmological parameters for the Discovery simulations were chosen based on the DESI Y1 BAO$+$CMB analysis.
The Discovery simulations ran on 960 nodes of Aurora, a GPU-accelerated exascale supercomputer hosted at the Argonne Leadership Computing Facility. 
With the Hardware/Hybrid Accelerated Cosmology Code (HACC), the run time for each simulation was $\mathcal{O}$(2 days).
These simulations serve as a demonstration of a unique new capability to run high-resolution simulations of cosmological volume in $\sim$2 days, allowing for close-to-real-time investigations of new cosmological results.

With the Discovery simulations, we have examined the way in which the change from a \lcdm cosmology to a \wowacdm cosmology leads to differences in the matter power spectrum, halo mass function, and halo mass accretion rate.
We have also applied a forward model for in situ star formation to investigate how changes in cosmology manifest as changes in galaxy star formation histories.
Overall, differences between the two Discovery simulations are at the level of $\sim$5-10\% for the measurements that we examined in this work.
Though these differences are small, the Discovery simulations provide a useful testbed for alternative cosmological probes that may offer additional constraining power for distinguishing between \lcdm and \wowacdm cosmologies, such as higher-order summary statistics and observables in the nonlinear regime.

FOF halo catalogs from the Discovery simulations are publicly available for $z=1.0,0.5,$ and 0, and can be downloaded from the HACC Simulation Data Portal.

%%%%%%%%%%%%%%%%%%%%%%%%%%%%%%%%%%%%%%%%%%%%%%%%%%

\section*{Acknowledgements}
We would like to thank the anonymous referee for their helpful comments and suggestions, which have improved the quality of this paper.
We thank Salman Habib, Eve Kovacs, Patricia Larsen, Matthew Becker, and other members of the Argonne CPAC group, as well as Johannes Lange, for useful discussions related to this work. We also thank Silvio Rizzi for visualizing the Discovery simulations. Work done at Argonne was supported under the DOE contract DE-AC02-06CH11357. We gratefully acknowledge use of the Bebop, Improv, and Swing supercomputers in the Laboratory Computing Resource Center at Argonne National Laboratory. The Discovery simulations were generated using resources of the Argonne Leadership Computing Facility, a U.S. Department of Energy (DOE) Office of Science user facility at Argonne National Laboratory and is based on research supported by the U.S. DOE Office of Science-Advanced Scientific Computing Research Program. This work was partially supported by the OpenUniverse effort, which is funded by NASA under JPL Contract Task 70-711320, ‘Maximizing Science Exploitation of Simulated Cosmological Survey Data Across Surveys’. The project that gave rise to these results received the support of a fellowship from "la Caixa" Foundation (ID 100010434). The fellowship code is LCF/BQ/PI23/11970028.

\bibliographystyle{aasjournal}
\bibliography{bibliography}

%\appendix
%\input{appendices.tex}

\end{document}